\documentclass[nofootinbib,reprint,amsmath,amssymb,aps]{revtex4-1}
\usepackage{graphicx}
\usepackage{dcolumn}
\usepackage{bm}
\usepackage{bbold}
\usepackage{slashed}
\usepackage{soul}
\usepackage{times}
\usepackage{epstopdf}
\usepackage{epsfig}
\usepackage{graphicx}
\graphicspath{ {images/} }
\usepackage[titletoc, title]{appendix}
\usepackage[colorlinks,
            linkcolor=red,
            anchorcolor=blue,
            citecolor=green
            ]{hyperref}
\usepackage{enumitem}
\begin{document}
\title{Does chiral perturbation theory rule out  QCD-based solutions to the strong CP problem?}
\author{Thomas D. Cohen} 
\affiliation{Department of Physics and the Maryland Center for Fundamental Physics\\University of Maryland\\}

\date{\today}

\begin{abstract}
The conventional view is that a solution of the strong CP problem lies beyond QCD.  A strong argument supporting this view is that the chiral  expansion shows that observables  depend on theta (unless a quark mass is zero); this eliminates the possibility that theta is physically irrelevant and appears to necessitate an explanation beyond the standard model.  However,  scenarios that solve the strong CP problem exist that are consistent with  known chiral behavior;  in these, QCD becomes nonviable as a theory for nonzero theta.  Such scenarios appear to be compatible with lattice studies of the topological susceptibility.  
\end{abstract}

\maketitle

\section{Introduction}

The resolution\cite{PhysRevD.14.3432,'tHooft:1986nc}  of the axial $ U(1)$ puzzle\cite{Weinberg:1975ui}---the absence of a ninth pseudo-Goldstone boson---gave rise to  the strong CP problem.   The solution of the $U(1)$ problem requires both the nonconservation of the axial $U(1)$  (due to the anomaly)  and a nontrivial topological susceptibility (initially envisioned in terms of instantons\cite{PhysRevD.14.3432,'tHooft:1986nc}),  which prevents the nonconservation from being rotated away.  However, nontrivial topological effects also imply that a possible CP-violating  term in the QCD Lagrangian (previously ignored as a total derivative) has physical consequences\cite{Callan:1976je,Jackiw:1976pf}.  The coefficient of it is $\overline{\theta}$  (after a chiral rotation to ensure a real mass matrix), which can take any value from $-\pi$ to $\pi$.  However, empirically it is exceptionally close to zero.  The exceedingly small value of the neutron electric dipole(the current bounds are  $-3.2 \times 10^{-26} \, e{\rm-cm} < d_n < 2.8 \times 10^{-26} \, e{\rm-cm}  $ (90\% C.L.) \cite{PhysRevD.92.092003} ) led to an  estimate\cite{RevModPhys.82.557} that $|\overline{\theta}|  \lesssim10^{-11} $.         This is an apparent violation  of Gell-Mann's famous dictum that particle physics is totalitarian in the sense that what is not forbidden is mandatory\cite{Gell-Mann:1956iqa}.  Why should a parameter not forbidden by symmetry be essentially zero?  This is the strong CP problem.  It  does not appear to be resolvable by anthropic arguments;  standard analysis suggests that the universe would not be substantially different than even if $\overline{\theta}$ were many orders orders of magnitude bigger than the current bounds\cite{Dine:2015xga}. 

The strong CP problem has motivated scenarios for beyond-the-standard-model (BSM) physics (for a review see ref.~\cite{RevModPhys.82.557}).   Indeed, new BSM solutions have been proposed in  the past few years \cite{Hook:2014cda,D'Agnolo:2015uta}.  The most influential class of  BSM proposals have been variations on the (approximate) symmetry of Peccei and Quinn (PQ) \cite{Peccei:1977hh,Peccei:1977ur} which, implies the existence of a psuedo-Goldstone boson, the axion\cite{Weinberg:1977ma,Wilczek:1977pj}.  Such an explanation is attractive in an additional way: the axion could be the dark matter which has long been known to exist from astrophysical evidence\cite{Bertone:2016nfn}.  However, major experimental efforts to detect the axion in a variety of ways (for a review see \cite{Graham:2015ouw}) have been unsuccessful to date.   It remains unclear if any of the proposed BSM solutions to the strong CP problem is correct.    Perhaps it is time to reexamine the possibility that the solution of the problem could lie within QCD itself. Scenarios with $m_u=0$ are not  QCD-based in this context; they require a BSM explanation of  why $m_u=0$. In any case, $m_u=0$ scenarios  are  ruled out by comparisons of current lattice studies to data\cite{Aoki:2016frl,Aoki:2013ldr,Fodor:2016bgu} which yield an up quark mass that differs from zero by more than 20 standard deviations.

Of course, there are very good reasons to reject a solution arising from QCD.  One of the strongest of these is the calculable and nonzero  $\overline{\theta}$ dependence of physical observables in an apparently reliable expansion scheme: chiral perturbation theory ($\chi$PT)\cite{Shifman:1979if,Brower:2003yx,Mao:2009sy,Aoki:2009mx,Guo:2015oxa}.  Superficially, the results of $\chi$PT by itself appear to exclude the possibility of a QCD-based solution.     However, it is worth recalling that systematic expansions can mislead.  Consider topological properties through the lens of a strict $N_c^{-1}$ expansion where $N_c$ is the number of colors\cite{Witten:1979vv,Veneziano:1979ec,DiVecchia:1980yfw}. This systematic expansion suggests that the topological susceptibility $\chi_T$ should be largely insensitive to the quark masses since the quark sector is formally down by $N_c^{-1}$.  However this is clearly wrong for very small quark masses.  The interplay between the chiral and $N_c^{-1}$  expansions is required to understand the correct behavior\cite{Witten:1979vv,Veneziano:1979ec,DiVecchia:1980yfw}.  The question addressed in this paper is whether the $\chi$PT-based analysis that apparently rule out  QCD-based solutions to the strong CP problem could similarly yield misleading results.   

Before addressing this question in detail, a word about notation.  CP violation can be introduced into QCD via two nominally distinct ways---either through the quark mass terms or via an explicit $\theta$ term.  However, physically these are not distinct: due to the anomaly, chiral rotations allows one to convert one type into the other without changing the physics.  The combination of parameters $\overline{\theta}$  determines the physical level of CP violation.   $\overline{\theta}$ is equal to the parameter specifying the $\theta$ term, after chiral rotations have been made that render all quark masses real and nonnegative.  Throughout this paper the formalism will be based on a representation of the theory that has all quark mass real and nonnegative.  Within this representation, there is no distinction between $\theta$ and $\overline{\theta}$.  Throughout this paper it will be denoted as $\overline{\theta}$ in order to make clear that it contains all of the CP violation in the theory.

\section{Conditions for a QCD-based solutions to the strong CP problem \label{conds}}

The perspective advocated here is that while $\chi$PT  presents severe challenges to QCD-based solutions, it appears possible to reconcile what is known  reliably about the chiral expansion with a solution to the strong CP problem within QCD. However such scenarios require that QCD behaves in a very  unexpected way in order to evade the strong constraints imposed by $\chi$PT.   Moreover, unlike the breakdown of the $1/N_c$ approximation when quark masses are small, there is no theoretical motivation for why QCD should behave this way---other than the need to solve the strong CP problem. By illustrating how peculiar QCD needs to  be to solve the problem by itself, this paper could simply be read as additional evidence for a BSM explanation.     However, given the importance of solving  the strong CP problem, it is important not to exclude possible solutions even they require unexpected behavior.

The most natural solution within QCD would be if $\overline{\theta}$ does not affect physical observables, despite the existence of nontrivial topological sectors.  Shifman, Vainshtein and Zakharov (SVZ) \cite{Shifman:1979if} considered  and rejected this possibility long ago on the basis of $\chi$PT; they showed that the anomaly plus the lack of a massless $\eta'$ in the chiral limit implies nonzero calculable CP violating observables at leading nontrivial order in $\theta$ and the quark masses.   Examples considered by SVZ included the amplitude for $\eta \rightarrow 2 \pi$ and the expectation value of the topological charge density, $ q(x) \equiv { g^2 F \tilde{F}(x)} /{32 \pi^2}$ at first order in $\overline{\theta}$.

For the present purpose it is more efficient to apply the logic  of Ref.\cite{Shifman:1979if} to the  topological susceptibility $\chi_T$ and higher-order cumulants of the topological charge defined by   \begin{align}
&c^{(2k)} \equiv \lim_{V \rightarrow \infty}\frac{(-1)^{k+1}\langle  Q^{2 k} \rangle_c }{V}  =\label{higherchi}\\
& \lim_{V \rightarrow \infty}\frac{ (-1)^{k+1}  \langle  Q^{2 k} \rangle - k(2k-1) \langle Q^{2 k-2} \rangle  \langle  Q^{2}  \rangle  + ...   }{V} \nonumber
\end{align}
(where the brackets indicate vacuum expectation value for $\overline{\theta}=0$, $Q \equiv \int {\rm d}^4 x  \, q(x)$, $V$ is the volume of (Euclidean) space-time,  $k$ are positive integers, the subscript $c$ indicates connected part and odd susceptibilities vanish due to CP)  are nonzero and calculable and $\chi_T \equiv c^{(2)}$.   To leading order in $\chi$PT they are given by 
\begin{equation}
\begin{split}
c^{(2k)}& =\left .\frac{-d^{2k} \left(m_\pi^2  f_\pi^2  \cos \left (\frac{\overline{\theta}}{2} \right )\sqrt{1 + \delta^2 \tan^2\left (\frac{\overline{\theta}}{2} \right )  } 
\right)} {d \overline{\theta}^{2k}} \right |_{\overline{\theta}=0}\\  & + {\cal O}\left ({m_\pi^4} \right )  \;\;\;\; {\rm with} \; \;\;\delta\equiv\frac{m_d-m_u}{m_d+m_u}\label{T2kch}
\end{split}\end{equation}
where  $f_\pi$ is the pion decay constant.
As stressed by SVZ,   Eq. (\ref{T2kch}) relies only on the anomaly, a standard chiral expansion which, for simplicity, only takes $m_u$ and $m_d$ as being small---to keep expressions concise QCD with two light flavors will be discussed here---and the absence of a massless $\eta'$ in the chiral limit.  All expressions are in Euclidean space.

 This paper assumes that $\chi$PT accurately reproduces all of the topological cumulants.  More precisely,  for each$c^{(2k)}$, at any given nontrivial order in $\chi$PT and with any fixed relative error tolerance, $\epsilon_R$, there exists a mass, $M$ such that whenever $m_\pi^2<M^2$, the $\chi PT$  value is within $\epsilon_R$ of the actual value.   This assumption is based on more than a natural prejudice that the chiral expansion, which accurately describes much of low energy QCD\cite{Scherer:2002tk}, should also work for the $c^{(2k)}$.   It is  also based on the fact that lattice studies have provided numerical evidence that $\chi$PT at low order accurately describes $\chi_T$ ( for state of the art results see refs.~\cite{Bonati:2015vqz,Aoki:2017paw}).     A key challenge of this paper is to reconcile this assumption with possible loopholes in reasoning based on $\chi$PT that exclude QCD-based solutions to the strong CP problem.

The challenge becomes clear from  standard functional arguments  which imply\cite{Guo:2015oxa} that the $c^{(2k)}$ can be written as 
\begin{equation} 
\begin{split}
c^{(2k)} &= \left .\frac{d^{2k} \epsilon( \overline{\theta})}{d \overline{\theta}^{2k}} \right |_{\overline{\theta} =0} \; {\rm with} \\ \epsilon\left (\overline {\theta} \right )  &\equiv -\lim_{V \rightarrow \infty} \frac{\log \left( 
\sum_Q  Z_Q(V) e^{i Q \overline{\theta}} \right)}{V}
\end{split}
\label{epsdef}\end{equation}
  where $\epsilon( \overline{\theta})$ is the energy density of the vacuum as a function of $\overline{\theta}$, 
$V$ is the Euclidean space-time volume and $Z_Q(V)$ is the partition function for a fixed topological sector.  Together with Eq.(\ref{T2kch}) this  implies that the energy density is given by 
\begin{align}
&\epsilon\left (\overline{\theta} \right )  -\epsilon(0)  = \label{eps} \\
&m_\pi^2  f_\pi^2 \left (1 -  \cos \left (\frac{\overline{\theta}}{2} \right )\sqrt{1 + \delta^2 \tan^2\left (\frac{\overline{\theta}}{2} \right )  } \right ) + {\cal O}(m_\pi^4). \nonumber
\end{align}
Clearly  the topological charge density, $q=\epsilon'\left (\overline{\theta} \right )$, a physically relevant observable, is nonzero and $\overline{\theta}$-dependent.  Moreover, it seems inconceivable that higher-order terms in the chiral expansion could act to exactly cancel out the dependence from the lowest order term.    This  provides a compelling reason to exclude QCD-based solutions to the strong CP problem that rely on all physical observables being independent of $\overline{\theta}$.

However, there is a loophole in this reasoning: QCD itself might solve that strong CP problem without requiring  physical observables to be independent of $\overline{\theta}$.   This would happen if some aspect of QCD renders the theory nonviable unless $\overline{\theta} \, ({\rm mod} \, 2 \pi)=0 $, even though $\overline{\theta}$ {\it apparently} affects physical observables.   Scenarios of this kind are analogous to triviality arguments for $\phi^4$ theory in 3+1 dimensions where  the value of the interaction term, which clearly would have a physical effect on the dynamics and is allowable by perturbative power counting, is required to be zero since otherwise the theory would not be ultraviolet complete. (The renormalization group using a perturbative kernel is suggestive of triviality; lattice studies are consistent with this expectation \cite{Luscher:1987ay,Luscher:1987ek}.  For a state of the art calculation see Ref. \cite{Korzec:2015pma}.)  
In a solution based on nonviability of QCD away from $\overline{\theta} \, ({\rm mod} 2 \pi)=0 $, some currently unrecognized feature of the theory (analogous to  the need for ultraviolet completeness) requires  $\overline{\theta}$ to be zero.  There were proposals along this general line in the past, for example Refs.~\cite{Schierholz:1994pb,Wu:1984bi}).  Scenarios such as these are highly speculative and hard to either verify or exclude.  In any case,  reconciling them with the SVZ-type analysis---which yield physical results at $\overline{\theta} \ne 0$ based on apparently reliable assumptions---appears to be problematic.  

Scenarios with QCD nonviable  for $\overline{\theta} \, ({\rm mod} 2 \pi) \ne 0$ have the function $\epsilon\left(\overline{\theta} \right )$  ill-defined when $\overline{\theta} \, ({\rm mod} 2 \pi)\ne 0$.  $\chi$PT  at any finite order yields an explicit and well-defined function for $\epsilon\left(\overline{\theta} \right )$,  presenting a challenge to such scenarios.  Moreover, an assumption underlying this paper, that the $c^{( 2k)}$---even derivatives of $\epsilon\left(\overline{\theta} \right )$ with respect to $\overline{\theta}$ (evaluated at $\overline{\theta}=0$)---are nonzero and accurately given in $\chi$PT makes this challenge acute.   Superficially, this seems to be an insurmountable challenge to such scenarios---how can a function that only exists at isolated points, have well-defined derivatives? Lattice calculations yield a nonzero susceptibility and appear to confirm $\chi$PT.  However, despite this challenge, scenarios of this type need not be ruled out due to $\chi$PT.    

This can happen, in principle, if $\epsilon\left(\overline{\theta} \right )$, while undefined for real $\overline{\theta}$ away from $\overline{\theta} \, ({\rm mod} 2 \pi)=0$, is never-the-less well-defined for  imaginary $\overline{\theta}$.  
Of course physically $\overline{\theta}$ must real due to unitarity.  However, mathematically additional information about  $\epsilon\left(\overline{\theta} \right )$  can be obtained by extending  Eq.~(\ref{epsdef}) to complex $\overline{\theta}$\cite{Cai:2016eot}.     The same logic based on generating functions that gave  in Eq.~(\ref{epsdef}) also yields $c^{\rm 2k}= (-1)^k \left . \frac{d^{2 k} \epsilon(\overline{i\theta})}{d (\overline{\theta})^{2k}}\right |_{\overline{\theta}=0}$.  Moreover, there is a caveat to  Eq.~(\ref{epsdef}): it holds,  only if the sum on $Q$ and infinite volume limit in the definition of $\epsilon( \overline{\theta})$ converge to a finite value for real $ \overline{\theta}$ in some finite neighborhood around $ \overline{\theta}=0$.   Suppose that it does not so that $\epsilon\left(\overline{\theta} \right )$ is ill-defined for real $\overline{\theta}$ with  $\overline{\theta} \, ({\rm mod} 2 \pi) \ne 0$, but  remains well-defined for imaginary $\epsilon\left(\overline{\theta} \right )$.  In such cases, it becomes possible that $\chi$PT could  accurately describes  $c^{(2k)}= (-1)^k \left . \frac{d^{2 k} \epsilon(\overline{i\theta})}{d (\overline{\theta})^{2k}}\right |_{\overline{\theta}=0}$.  The significant point is that the existence of nonzero topological susceptibilities need not require that $\epsilon( \overline{\theta})$ to be well defined on the real axis away from $\overline{\theta} \, ({\rm mod} 2 \pi)=0$; it is sufficient for it to be well-defined on the imaginary axis.  In the remainder of this paper, it will be assumed that this occurs.  Such scenarios depend on an obstruction to analytically continuing the functional form of the energy density from imaginary $\overline{\theta}$ to the real axis and this implies that the point $\overline{\theta}$ is nonanalytic. 

The challenges posed by $\chi$PT to a  solution to the strong CP problem within  QCD can be met provided $\epsilon(\overline{\theta})$ and  the $c^{(2k)}$ satisfy the following conditions: 
 \begin{enumerate}[label=(\roman*)]
\item $\epsilon(\overline{\theta})$ is finite and real for $\overline{\theta}$  real with $\overline{\theta} \, ({\rm mod} \, 2 \pi) =0$ and for $\overline{\theta}$ purely imaginary, \label{cond1}
\item   $\epsilon(\overline{\theta})$ is ill-defined corresponding to a nonviable physical theory when  $\overline{\theta} \, ({\rm mod} \, 2 \pi)  \ne 0$  with $\overline{\theta}$ real. \label{cond2}
\item $\epsilon(\overline{\theta})$ is an analytic function for  $\overline{\theta}  \ne 0$  with $\overline{\theta}$ purely imaginary. \label{cond3}
\item $\epsilon(i \overline{\theta})$ is not analytic at $\overline{\theta} =0$. \label{cond4}
 \item All derivatives of  $\epsilon(i \overline{\theta})$ with respect $\overline{\theta}$ are finite and well-defined at  $\overline{\theta} =0$, despite the nonanalyticity at zero of condition \ref{cond4}.   \label{cond5}
\item  The topological cumulants are well-defined, finite and given by $c^{(2k)}= (-1)^k \left . \frac{d^{2 k} \epsilon(\overline{i\theta})}{d (\overline{\theta})^{2k}}\right |_{\overline{\theta}=0}$\label{cond6}.
\item  For any $c^{(2k)}$ and any given nontrivial order  in $\chi$PT with any fixed relative error tolerance , $\epsilon_R$, there exists a mass, $M$ such that whenever $m_\pi^2<M^2$, $\chi PT$  accurately reproduces $c^{(2k)}$ to within $\epsilon_R$. \label{cond7}
\item The chiral expansion for   $\epsilon(\overline{\theta})$ when $\overline{\theta}$ is purely imaginary is asymptotic; for any fixed order, any imaginary $\overline{\theta}$ and any fixed relative error tolerance, $\epsilon_R$,  there exists an $M$ such that whenever $m_\pi^2<M^2$, $\chi$PT accurately  reproduces $\epsilon(\overline{\theta})$ to within $\epsilon_R$.     \label{cond8}
\item The nonanalyticity of   $\epsilon(i \overline{\theta})$ at $\overline{\theta} =0$ of condition \ref{cond4} is due to contributions that are subleading to all orders in $\chi$PT.  \label{cond9}
\end{enumerate}

Conditions \ref{cond1}  and \ref{cond3} encode the expected behavior when $\overline{\theta}$ is imaginary that lead to well-defined topological susceptibilities, while  condition \ref{cond2} is the heart of scenarios  that make QCD nonviable for  real $\overline{\theta}$ away from $\overline{\theta} \, ({\rm mod} 2 \pi)=0$.   Conditions  \ref{cond6} and \ref{cond7} place severe constraints on such scenarios due to $\chi$PT.  Conditions \ref{cond4} and \ref{cond8} play essential roles in evading such constraints: Condition \ref{cond4} reconciles conditions \ref{cond2} and \ref{cond3},   while condition \ref{cond9}  reconciles conditions \ref{cond2}, \ref{cond4},\ref{cond3} and \ref{cond8}.

\section{An illustrative example \label{IE}}

Consider the following illustrative functional form which illustrates the type of function that satisfies all 9 conditions: 
\begin{subequations}
\begin{align}
&\epsilon^{\rm illus}\left (\overline{\theta} \right )  - \epsilon(0) \label{intform}\\ 
& =  m_\pi^2 f_\pi^2 \,  \, g\left(\overline{\theta},\delta\right) \int_0^\infty d x \, e^{-x + x^2 \, \left (\frac{m_\pi^2 f_\pi^2}{\Lambda^4}  \right ) \, g\left(\overline{\theta},\delta\right)} \nonumber\\
& = \sum_{j=0}^\infty  \frac{(2j)! }{j!}  \Lambda^4  \left (\frac{m_\pi^2 f_\pi^2}{\Lambda^4}  \right )^{j+1}\, g\left(\overline{\theta},\delta\right)^{j+1} \label{epsexpand}\\
&=  \Lambda^4 \left ( \left (\frac{m_\pi^2 f_\pi^2 \, g\left(\overline{\theta},\delta\right) }{\Lambda^4}  \right )+ 2 \left (\frac{m_\pi^2 f_\pi^2 \, g\left(\overline{\theta},\delta\right) }{\Lambda^4}  \right )^2\, + \right .  \nonumber\\
& \left . 12  \left (\frac{m_\pi^2 f_\pi^2 \, g\left(\overline{\theta},\delta\right) }{\Lambda^4}  \right )^3 
+ 120  \left (\frac{m_\pi^2 f_\pi^2 \, g\left(\overline{\theta},\delta\right) }{\Lambda^4}  \right )^4 + \cdots\right ) \nonumber  \\
&{\rm with}  \;  \;  g\left(\overline{\theta},\delta\right) =1 -  \cos \left (\frac{\overline{\theta}}{2} \right )\sqrt{1 + \delta^2 \tan^2\left (\frac{\overline{\theta}}{2} \right )  }  \label{g}\\
 & = \frac{(1-\delta^2) \, \overline{\theta}^2}{8}  - \frac{(1+2 \delta^2-3\delta^4)\, \overline{\theta}^4 \,  }{384}  +  \cdots  \nonumber
 \end{align}
\end{subequations} 
where $\Lambda$ is a parameter with dimensions of mass.  

Equation~(\ref{epsexpand}), the  formal chiral expansion of  $\epsilon^{\rm illus}$, is asymptotic and valid when $\overline{\theta}$ is imaginary; at lowest order it yields Eq.~(\ref{eps}).  The form of $g\left(\overline{\theta},\delta\right)$ in Eq.~(\ref{g}) accounts for isospin violation and ensures that all $\overline{\theta}$ dependence vanishes when $m_u$ or $m_d$ is zero.  Note that  $g\left(\overline{\theta},\delta\right)$  has an important property: it is greater than zero when $\overline{\theta}$ is real and away from  $\overline{\theta} \, ({\rm mod} 2 \pi)=0 $, equal zero when $\overline{\theta} \, ({\rm mod} 2 \pi)=0$ and less than zero when $\overline{\theta}$ is imaginary.    From  the integral form of Eq. (\ref{intform}), this implies that energy density is divergent for real  $\overline{\theta} \, ({\rm mod} \, 2 \pi) \ne 0$ but convergent for imaginary  $\overline{\theta}$; thus,
conditions \ref{cond1} and \ref{cond2} hold. One might hope that despite the nonanalytic behavior at  $\overline{\theta}=0$, one could analytically continue the  $\epsilon^{\rm illus}$ from the imaginary axis to real $\overline{\theta}$.  One can; but the function so obtained  is multibranched and all branchs have an imaginary part---indicating that they are not physically viable.

It  is straightforward to verify that conditions \ref{cond1}-\ref{cond9} hold for $\epsilon^{\rm illus}$.  The numerical coefficients in the chiral expansion of Eq.~ (\ref{epsexpand}) explain why: the coefficient of  the $j^{\rm th}$ term is $\frac{(2j)!}{j!}$, which grows faster than any power law in $j$.  Thus the radius of convergence for the chiral expansion is strictly zero.  Moreover, the series is actually in $m_\pi^2  g\left(\overline{\theta},\delta\right)$ where  $g\left(\overline{\theta},\delta\right)$ in  a series in $\overline{\theta}$ begins at order $\overline{\theta}^2$.  Thus, the radius of convergence of an expansion in $\overline{\theta}$ is also zero and the the point $\overline{\theta}=0$ is nonanalytic.   The rapid growth in these coefficients implies that while  $\chi$PT at low order can accurately describe the $c^{(2k)}$ (as required by condition \ref{cond7}),  the value of $m_\pi$ for which $\chi$PT at fixed order is accurate rapidly drops with $k$.  

If $\epsilon\left (\overline{\theta} \right )$ in QCD were given by $\epsilon^{\rm illus}\left (\overline{\theta} \right )$, the strong CP problem would be solved.    Of course, in QCD $\epsilon\left (\overline{\theta} \right ) \ne \epsilon^{\rm illus}\left (\overline{\theta} \right )$, which was given simply to illustrate that functions satisfying all the conditions exist.  $\epsilon^{\rm illus}\left (\overline{\theta} \right )$ is consistent with a chiral langrangian given to all orders in $m_\pi^2$ but treated at tree level.  In practice, higher-order terms  in $\chi$PT  will develop chiral logarithms due to infrared behavior in loops along with the powers of $m_\pi^2$ associated with tree-level terms \cite{Brower:2003yx,Mao:2009sy,Aoki:2009mx,Guo:2015oxa}.  However,  these chiral logs are entirely fixed by lower-order terms in the theory and should not affect whether or not the conditions--are satisfied.  The critical issue for this is whether the coefficients of higher-order terms in the chiral expansion of $\epsilon\left (\overline{\theta} \right )$ grow sufficiently rapidly;  at this stage  we do not know whether or not they do in QCD.

This paper has focused on $\epsilon(\overline{\theta})$.  However, it should be clear that if scenarios of this sort were valid, other observables such as the $\eta \rightarrow 2 \pi$ amplitude considered by SVZ\cite{Shifman:1979if} would be expected to have analogous behavior.  The matrix elements for these observables would be finite and describable  in $\chi$PT for imaginary $\overline{\theta}$ but would become ill-defined for real  $\overline{\theta} \, ({\rm mod} 2 \pi) \ne 0$, the function would be nonanalytic at  $\overline{\theta}=0$ with the nonanalyticity  subleading to all orders in $\chi$PT.

\section{Discussion}

The central argument of this paper is that there is a loophole in the logic that $\chi$PT excludes QCD-based solutions to the strong CP problem.   Of course, there remain very strong reasons to doubt that QCD itself can be the solution of the strong CP problem.    One of these is simply that the loophole requires QCD to behave in a very surprising manner that differs from the familiar ways we expect quantum field theories to behave.   Such behavior is radically differently from any quantum field theory that the community has seen over the decades.  A second reason is that the motivation for considering the conditions proposed here is entirely phenomenological; they were proposed solely to ensure that  $\overline{\theta} \, ({\rm mod} 2 \pi) \ne 0$  without violating established properties of QCD.  Accordingly, there is no underlying theoretical basis for expecting QCD to satisfy them. The illustrative model of Sect. \ref{IE} illuminates this.  While the illustrative model satisfies conditions \ref{cond1}-\ref{cond9} and thereby demonstrating that forms that do so are not excluded mathematically, it is also contrived.  The model did not emerge as a natural outgrowth of any theoretically-motivated mechanism, rather it was essentially reverse engineered for the sole purpose of satisfying the conditions. The upshot of this, is that it is hard to see why a functional  form satisfying the conditions should emerge from QCD.

 On the other hand, there are no easy solutions to the strong CP problem.  Solutions require one to conjecture entire new sectors of BSM physics---a major intellectual leap---or to  conjecture that the standard model behaves in a very unfamiliar way---which is also a major intellectual leap.  Thus, it seems sensible to explore the possibility of all possible solutions.

It is important to stress the way in which this class of scenarios solves the strong CP problem.  In these  scenarios the theory is only viable as a physical theory for values of $\overline{\theta}$ satisfying  $\cos\left (\overline{\theta} \right ) = 1$.  At first glance, this might appear to be inconsistent with $\epsilon(\overline{\theta})$ being well-defined and analytic when  $\overline{\theta}$ is purely imaginary.  Indeed, one might worry that the existence of well-defined function on the imaginary axis means that the theory allows CP violating physics provided that one takes $\overline{\theta}$ to be imaginary.  However, such concerns are misplaced.  It is important to distinguish between the theory as a mathematical object and one describing physics.  Physically, the the theory is only sensible when  $\overline{\theta}$ is real.  If it is not, the theory is not unitarity and is not an acceptable description of nature.  On the other hand, one can define the path integral for the theory  mathematically regardless of the phase of  $\overline{\theta}$ (assuming that it converges appropriately) even when this renders the theory unphysical.   Thus if this class of scenario is correct, then the strong CP problem is solved: the only physically viable value of $\overline{\theta}$ in the theory has  $\cos\left (\overline{\theta} \right ) = 1$ and no CP violation.  This remains true despite the fact that theory is mathematically sensible for unphysical imaginary values of $\overline{\theta}$.

Clearly solutions to the strong CP problem that exploit the loophole noted in this work are very different from axion solutions.  In axion models,   $\overline{\theta}$ effectively becomes dynamical; its value is free to respond to the environment.  The strong CP problem is solved since the effective potential  for $\overline{\theta}$ has a  minimum at $\cos \left( \overline{\theta} \right)=1$; if a region of space had $\cos\left( \overline{\theta} \right ) \ne 1$, it would not be in a stable equilibrium and would slide toward the minimum.  In contrast, solutions of the sort considered here, $\overline{\theta}$ does not adjust itself to a value where CP violation vanishes.   Rather $\overline{\theta}$ is not dynamical, but a fixed value.  The solution requires that for reasons currently not understood---but also not completely ruled out given our current state of knowledge---QCD does not exist as a viable physical theory unless  $\cos \left( \overline{\theta} \right)=1$; thus, if the standard model  contains QCD it must have  $\overline{\theta} \, ({\rm mod} 2 \pi) = 0$.

Ideally one should be able to verify or exclude this class of solution---or at least develop evidence in favor or opposed to it.   A key problem with trying to rule out scenarios of the sort considered here is  their phenomenological (as opposed to  theoretical) motivation.   This lack of an underlying theoretical picture means that one cannot rule out this class of scenario by undermining the underlying theoretical assumptions.   

Never-the-less, one obvious future research direction is to see whether there are reliable theoretical arguments that can close the loophole consider in this paper.  In essence that would mean a ``no-go'' theorem that would rule out this class of solutions to the strong CP problem.   In this context, the interplay between perturbative and nonperturbative aspects of a chiral expansion suggest analysis along using ideas of resurgence (for a review of the state of the art see\cite{Aniceto:2018bis,Dunne:2016nmc}) might shed light on the issue.

If the loophole cannot be closed using reliable theoretical arguments,  other ways of obtaining evidence in support or against solutions of this type should be considered.  Clearly, the most straightforward way to exclude these scenarios would be via a direct and reliable calculation of $\epsilon \left(\overline{\theta}\right)$.  If one could show that the $\epsilon \left(\overline{\theta}\right)$ is well-defined and calculable in QCD away from $\cos\left( \overline{\theta} \right)=1$,  the scenario is ruled out.   Unfortunately the only known systematic reliable numerical method for computing nonperturbative observables directly from QCD is via Euclidean-space lattice calculations and  direct lattice calculations of   $\epsilon \left(\overline{\theta}\right)$ are well beyond the state of the art due to a sign problem\cite{Cai:2016eot}.   Accordingly, it is necessary to consider what kind of indirect evidence one can obtain.

As it happens, if such scenarios were correct, Monte Carlo calculations in Euclidean space without exponentially bad sign problems could, in principle,  provide compelling, if indirect, evidence for them.  However there are severe practical limitations to such calculations for QCD.   These practical limitations will almost certainly make any such evidence obtainable in the foreseeable future quite indirect.  Still, even indirect evidence for such scenarios would be very significant.   

The basic reason why lattice calculations could,  in principle,  provide indirect support for such scenarios comes from conditions \ref{cond1}-\ref{cond6}, which do not depend in detail on chiral properties.  These provide a distinctive signature.   Consider the function $\epsilon \left (\overline{\theta}\right)$ along the imaginary $\overline{\theta}$ axis where  it is well defined in the scenarios considered here.  The conditions imply that at $\overline{\theta}=0$ all derivatives of  $\epsilon \left(i \overline{\theta}\right)$ are finite and well defined but the function is nonanalytic.  Thus, if one were to write as $\epsilon \left(i \overline{\theta}\right)$ as a Taylor expansion around some point in the complex plane, its radius of convergence would approach zero as that point approaches $\overline{\theta}=0$.   

Given that $\epsilon \left (\overline{\theta}\right)$ is an even function and that up to a k-dependent sign , the topological cumulants, $c^{(2k)}$ are given by the $(2k)^{\rm th}$ derivative of  $\epsilon \left(i \overline{\theta}\right)$ with respect to $\overline{\theta}=0$, it follows from the ratio test that, $R$, the radius of convergence  of a Taylor expansion for $\epsilon \left (\overline{\theta}\right)$ in general is given  by
\begin{equation}
R=\lim_{n \rightarrow \infty} \sqrt{\frac{ \frac{|c^{(2n) }|}{(2n)!}}{\frac{ |c^{(2 n+2)}|}{(2n+2)!}} }  \label{radius} \end{equation}
provided the limit exists.  This means that if
\begin{equation}
\lim_{n \rightarrow \infty} f_n   \rightarrow \infty \; \; \; {\rm where} \; \; f_n  \equiv \frac{ |c^{(2 n+2)}|}{(n+1)(n+2)|c^{(2n)}|} \; ,
\label{ratio}
\end{equation}
 the radius of convergence is zero, a necessary condition for this class of scenario to be valid.
Provided that such behavior could be documented for QCD, one would have strong evidence in favor of such scenarios.

Since topological cumulants are calculable  without a sign problem (either by directly computing fluctuations in the topological charge or by computing numerical derivatives of  $\epsilon \left(i \overline{\theta}\right)$), one could look for indications that  in QCD  $f_n$ grows without bound as $n \rightarrow \infty$ (as is required by scenarios of the sort considered here).  Such studies are necessarily indirect: lattice calculations can only provide information about  a finite number of topological cumulants.  Thus one cannot determine that  $f_n$ is actually diverging or not; the best one can do is see whether it {\it appears} to be growing in a manner consistent with divergence.  Still a convincing indication of this sort might be regarded as compelling evidence for such scenarios.

\begin{figure*}[t]
\centering
\includegraphics[width=1. \textwidth]{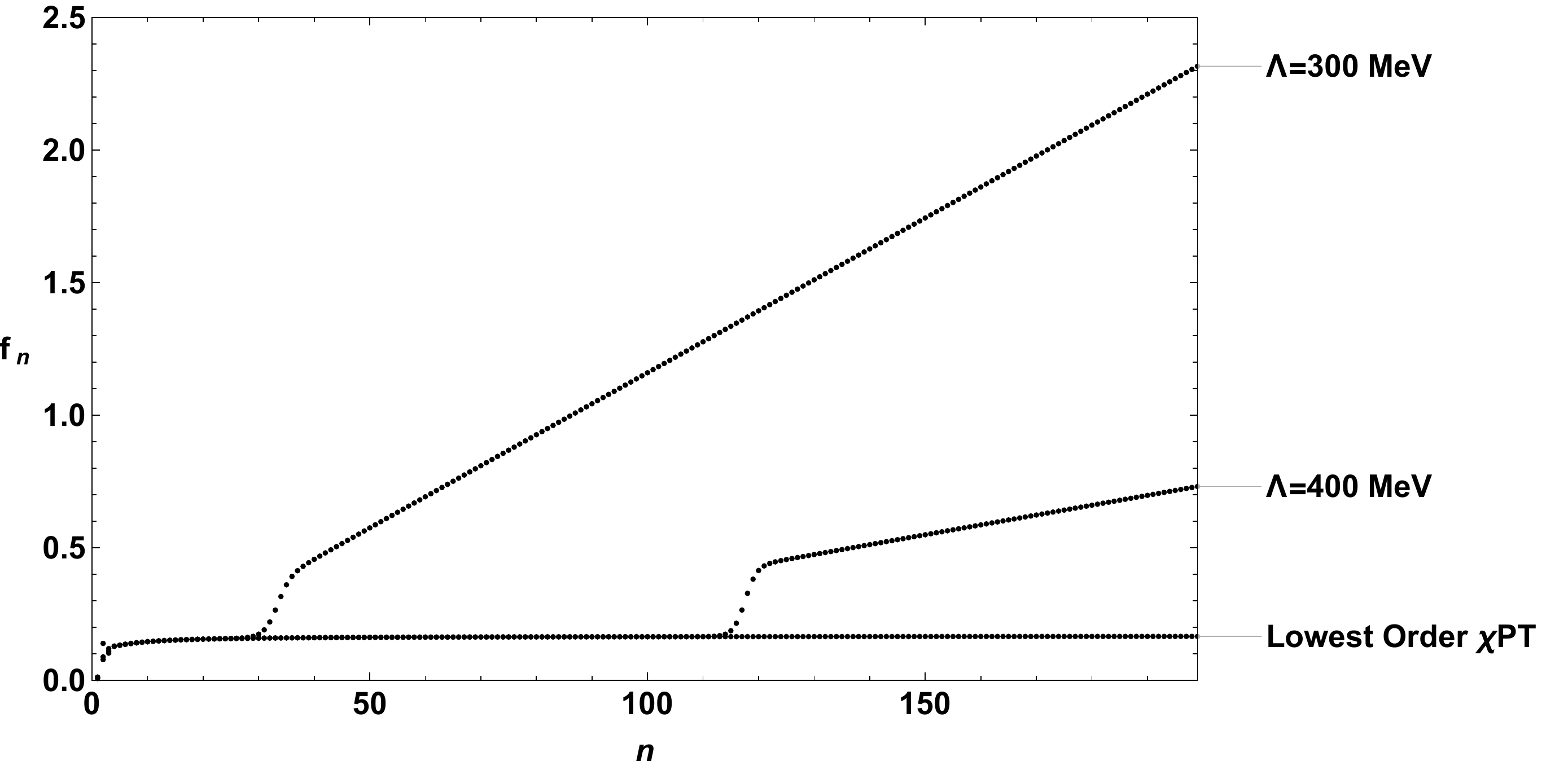}
    \caption{$f_n \equiv \frac{ |c^{(2 n+2)}|}{(n+1)(n+2)|c^{(2n) }|}$ for $\epsilon^{\rm illus}\left(\overline{\theta}\right)$, the illustrative model of Sec. \ref{IE}.  The $c^{(2 n)}$ are topological cumulants. Two different values of the parameter $\Lambda$ are shown. For comparison the same ratio is also given for lowest order chiral perturbation theory.   If this ratio diverges as $n \rightarrow \infty$, then $\epsilon\left(i \overline{\theta}\right)$  is nonanalytic at $\overline{\theta}=0$.}
    \label{fig:toy}
\end{figure*}

To see how in principle this could work,   suppose for the sake of illustration that $\epsilon \left(i \overline{\theta}\right)$ in QCD is given exactly by $\epsilon^{\rm illus}\left(i \overline{\theta}\right)$, the illustrative model of Sec.~\ref{IE} and that lattice calculations determining numerous topological cumulants with negligible error were tractable.   In Fig.~\ref{fig:toy} the ratio $f_n$ defined in Eq.~(\ref{ratio}) is plotted as a function of $n$ for this  model with two different values of the parameter $\Lambda$, 300 MeV and 400 MeV; for comparison this ratio is also given for lowest order chiral perturbation theory.    The ratio in lowest-order chiral perturbation theory saturates at a finite value, as one expects when $\epsilon \left(i \overline{\theta}\right)$ has a non-zero radius of convergence.  In contrast, for the toy model with either $\Lambda =$ 300 MeV or  $\Lambda =$ 400 MeV, the ratio quite clearly appears to be growing    linearly  with $n$ asymptotically.   To the extent that the behavior is indicative of the true asymptotic behavior (as is the case for the toy model), one can conclude that the Taylor series for  $\epsilon \left(i \overline{\theta}\right)$ around zero has a zero radius of convergence as required by the scenarios considered here.  The behavior seen in  Fig.~\ref{fig:toy} is quite dramatic.  If one could obtain lattice data of this sort for QCD and found that it had this behavior one would have compelling, if somewhat indirect, evidence for a scenario of the sort considered here.

Unfortunately, Fig.~\ref{fig:toy} also indicates practical problems in implementing such an approach that render it intractable for QCD.  Note that for low $n$,  the data for  $f_n$ does not even hint that it will ultimately grow linearly with $n$ for asymptotically large $n$.  Rather,  up to a fairly large value of $n$, the ratio appears to be saturating to a finite value as one would expect with an analytic function around  $\overline{\theta}=0$---falsely suggesting that this functional forms does not satisfy the conditions.   For the model with  $\Lambda =300$ MeV,  the first noticeable hint that the ratio is not  saturating is around  $n=30$ (requiring a calculation of  topological cumulants up to $c^{(62)}$);    For the model with  $\Lambda =400$ MeV, the first noticeable hint that it is not saturating is around  $n=110$  (requiring a calculation of  topological cumulants up to $c^{(222)}$).   

The sharp onset at some large value of $n$ of behavior inconsistent with saturation  makes verifying such a scenario for QCD highly problematic even assuming QCD behaved according to this class of scenario.  Suppose hypothetically this class of scenario were correct for QCD and  moreover algorithms for the determination of topological cumulants of QCD  advanced to the point where they could be calculated with high accuracy up to $c^{(40)}$ (which corresponds to $n=19$).    Even with this large number of topological cumulants, it could easily still  be the case one could have no numerical evidence for the scenario.  It is not implausible given the toy model one might need up accurate cumulants up through $c^{(100)}$ ({\it i.e.} $n=49$) or some larger value before there was compelling evidence for the scenario.  

In any case, a reliable and accurate calculation of topological cumulants through $c^{(40)}$ will almost certainly be well beyond our reach for QCD for the foreseeable future---unless some radically improved algorithm is found.  Although topological cumulants for fixed $n$ are calculable on a Euclidean lattice studies without exponential sign problems, with all current approaches, the difficulty in computing them grows with the order of the cumulant.  If one computes the $c^{(2 n)}$  via fluctuations in the topological charge the signal to noise worsens as $n$ increases.  if one attempts to extract them as numerical derivatives of $\epsilon \left(i \overline{\theta}\right)$ one is faced with the need for computing $\epsilon \left(i \overline{\theta}\right)$ with increasing accuracy to accurately get higher derivatives; this is particularly difficult given that the function is being evaluated numerically via Monte Carlo methods.  In point of fact, even for the computationally much simpler case of Yang-Mills theory, only a few topological cumulants have been computed or even bounded (see, for example ref.~\cite{Bonati:2015sqt}).

The qualitative behavior of Fig.~\ref{fig:toy} in which $f_n$ appears to be saturating with $n$ until at some comparatively large value of $n$ there is a sharp onset of behavior inconsistent with saturation is easy to understand.   Condition \ref{cond9} implies that the nonanalyticity of   $\epsilon(i \overline{\theta})$ at $\overline{\theta} =0$  is due to contributions that are subleading to all orders in $\chi$PT.   Thus comes about because there are contributions to $\frac{c^{2n }}{(2 n)!}$  that simultaneously diminish with $n$ due to a chiral suppression that scales like $\left (\frac{m_\pi}{\Lambda'}\right )^{2 n}$  (where $\Lambda'$ is a typical hadronic scale) and, due to a numerical factor, grows with $n$  faster than a exponentially.  This behavior can easily be seen to occur in $\epsilon^{\rm illus}$, the toy model of  Sec.~\ref{IE}.  For small values of $n$, the chiral suppression overwhelms the numerical coefficient and these contributions  are swamped by contributions that are leading order in a chiral expansion.  However the numerical coefficients grow faster with $n$ than the diminution due to  chiral suppression.  Thus at some value of $n$, this contribution ceases to be negligible.  Moreover, the rapid growth of the numerical contribution, implies that once the term ceases to be  negligible, it rapidly becomes dominant.  Thus, one expects the sort of rapid onset of behavior incompatible with saturation seen in Fig.~\ref{fig:toy}.   

The value of $n$ where this change of behavior sets in, depends on the scale of the chiral suppression.  In the toy model this is fixed by the dimensionless combination $\frac{m_\pi^2 f_\pi^2 }{ \Lambda^4} $.  Thus larger values of $\Lambda$ in the model will lead to more chiral suppression and accordingly the onset of growth incompatible with saturation occurring at larger $n$.  This is seen in  Fig.~\ref{fig:toy} where for $\Lambda$=400 MeV, this onset is pushed out to the neighborhood of $n=110$, whereas it occurs around $n=30$ for $\Lambda$ =300 MeV.

One might worry that the prospects for directly detecting this sort of behavior in QCD may  be even more dire than suggested by Fig.~\ref{fig:toy}.  The two models used to illustrate the issue had  $\Lambda =300$ MeV and $\Lambda =400$ MeV which might be regarded as quite low scales.  The parameter $\Lambda$ controls the scale of the chiral suppression and one might assume that it should be taken to be naturally at a typical hadronic scale of order 1 GeV;  values $\Lambda$ at that scale would push the  onset of behavior incompatible with saturation in this model to extremely large values of $n$.   

On the other hand, it is hard to estimate the natural scale for $\Lambda$.   The size of the chiral suppression in the model is fixed by the combination $\left ( \frac{m_\pi^2 f_\pi^2 }{ \Lambda^4} \right )$ whose form was picked in order to have the $m_\pi^2 f_\pi^2$ structure match with the leading order $\chi PT$ result.  However, $f_\pi^2$ is numerically quite small on the scale of hadronic physics.   One could just as well have written this as $\frac{m_\pi^2}{\Lambda'^2}$ with $\Lambda' =\frac{\Lambda^2}{f_\pi}$.  The models with $\Lambda=$ 300 MeV  correspond to  $\Lambda'=$ 958 MeV  which may beregarded as naturally sized while $\Lambda=$ 400 MeV  corresponds to $\Lambda'=$  1720 MeV  which may regarded as large. The numerical value of the parameter one uses to parameterize the chiral suppression is a matter of bookkeeping as well physics.  Regardless of how natural the model parameters are, the model strongly suggests that even if the scenarios considered in this paper were correct, it seems extraordinarily unlikely that this would be revealed by lattice studies of the topological cumulants in the foreseeable future.

Fortunately one need completely not rule out the possibility of indirect evidence in support of such scenarios.  For example,  it is conceivable that  volume dependence could be used to obtain evidence for or against the type of QCD-based solution to the strong CP problem considered in this paper.  It has long been known\cite{Hansen:1990yg,Leutwyler:1992yt} that interplay between topology, the chiral limit and the infinite volume limit is subtle.  However, by exploiting known behaviors near the chiral limit one can make concrete predictions for topological behavior taking into account finite volume effects, for example in the so-called $\epsilon$-regime.  Unfortunately, one cannot just borrow these results here since the essence of the current scenarios is that the leading order chiral effects yield misleading results for the $\theta$ dependence.  Never-the-less, one might imagine that if the full interplay of topological,  finite volume and chiral effects were understood in the context of this class of scenario, then finite-volume studies might have distinctive behavior which might act as a signature.  If things were particularly fortuitous, it is possible that such volume dependent behavior signature might  turn out to be far more practical then directly computing numerous topological cumulants.  

However, since this class of scenario was proposed for essentially phenomenological reasons and lack a theoretically motivated detailed mechanism it is not currently possible to deduce from first principles how finite volume effects will alter the delicate interplay of chiral effects with analyticity of  behavior of $\epsilon(\overline{\theta})$ near $\theta=0$.  Thus, one cannot easily anticipate whether finite volume effects might prove to be a useful tool in discerning whether QCD obeys such a scenario. 

If the solution to the strong CP problem is due to a scenario of this type, there is another way to find evidence in its favor.  Recall that a principal reason that  the scenarios consider here may seen implausible {\it a priori}  is that after decades of experience with quantum field theory, the community has never encountered any theories that behave in such a manner.  Thus, a more general strategy would be to concentrate on quantum field theories that, while distinct from QCD, are more or less related and more tractable.  If one could see compelling evidence for behavior analogous to conditions \ref{cond1}-\ref{cond9} in such a theory that would greatly increase the plausibility of the scenario.  Indeed evidence for behavior analogous conditions \ref{cond1}-\ref{cond6} would be enough to substantially increase the plausibility; one can view conditions \ref{cond1}-\ref{cond6} as the critical underlying ones while conditions \ref{cond7}-\ref{cond9} could emerge naturally in QCD if the first six held, in order to reconcile chiral physics with the behavior of implied by  \ref{cond1}-\ref{cond6}.  

The behavior exhibited in Fig.~\ref{fig:toy} suggests one obvious line of attack.  Evidence that the ratio $f_n$ is not saturating to a finite value (as expected from models satisfying the first six conditions) requires far smaller values of $n$ for the model with $\Lambda=$ 300 MeV then the model with $\Lambda=$ 400 MeV.    The difference between the two models is that the chiral suppression is more pronounced for the $\Lambda=$ 400 MeV case.  Indeed, conditions \ref{cond6}-\ref{cond9} imply that as the chiral suppression grows, the value of $n$ needed before the onset of non-saturating behavior does as well.  Similarly as the chiral suppression shrinks, the value of $n$ needed before the onset of non-saturating behavior does as well.    In QCD, one can make the size of chiral suppression small, simply by increasing the value of the quark masses.

If QCD  behaves according to a scenario of this type,  one might hope, optimistically, that with sufficiently large quark masses, evidence $f_n$  grows with $n$ in a manner suggestive of behavior inconsistent with saturation might be obtainable for sufficiently small $n$ that such calculations might be tractable---at least in the foreseeable future.  

There are a number of caveats to this.  Clearly if such behavior is seen with large quark masses, it will not fully establish the scenario.  If chiral suppression play no role in the observation, the best one can test are conditions \ref{cond1}-\ref{cond6} and not conditions \ref{cond6}-\ref{cond9}, which depend on chiral symmetry.   Moreover, it is at least theoretically conceivable that scenarios of this sort only apply for a certain domain of quark masses and that this domain includes large quark masses but not realistic ones. This is a minor concern.  If, contrary to all previous experience with quantum field theory, QCD with large quark mass satisfies conditions \ref{cond1}-\ref{cond6} (providing a natural explanation for the strong CP problem), it becomes extremely plausible  given the empirical fact the CP violation in strong interactions is nonexistent or extremely small, that the same behavior would extend down to light quark masses and thereby requiring conditions \ref{cond6}-\ref{cond9}.  

A more significant concern given the limited number of topological cumulants that are likely to be accessible is the possibility of a ``false positive'' in which the  ratio $f_n$ is seen to be growing with $n$ for a few calculable small values of  $n$ in a manner suggesting that the ratio may not saturate, when in fact it does, but does so at slightly large values of $n$.   There is also the possibility of ``false negatives''.  One situation that could arise is that the behavior near the chiral limit play a central role in realizing such scenarios and that they only are realized in QCD for a domain in which the quark masses are sufficiently small.  If this were the case then calculations at large quark masses would miss the effect.   

There is also a  practical concern.  One is likely to be able to compute only a small number of topological cumulants.  It is quite possible that for those $n$ which are calculable  the  $f_n$ appears to be saturating with $n$  but ultimately there will be the onset of behavior where it increases without bound but this onset is in a regime beyond where the $c^{(2n)}$ are calculable---even if the quark masses are large.   

If one is attempting to render a study tractable by increasing the quark masses, it is reasonable to consider the extreme case where they go to infinity leaving a pure gauge theory.  Doing this not only eliminate chiral suppression altogether leading to the prospect of seeing signatures of the scenario at smaller values of $n$, it has the obvious practical advantage that calculations lacking a quark functional determinant are far more straightforward numerically and hence one can compute $c^{(2n)}$ coefficients to larger values of $n$.  Of course, removing quarks entirely from the problem makes the theory that much further from QCD and thus that harder to draw definitive conclusions.   Still, if nontrivial evidence that Yang-Mills theory appears to be consistent with conditions \ref{cond1}-\ref{cond6}, it would go a long way towards establishing a QCD-based solution to the strong CP problem: the fact that such conditions were satisfied for a theory related to QCD combined with the phenomenological fact that CP violation in QCD is either nonexistent or very small would make such a scenario  plausible,

While Yang-Mills theory has no chiral suppression, numerical challenges may emerge due to suppression in  $1/N_c$ ( where $N_c$ is the number of colors in the theory)---but fortunately these are much less severe than those due to chiral suppression.     Witten showed long ago\cite{Witten:1979vv} that standard $N_c$, counting rules  imply that as the large $N_c$ limit is approached, the topological cumulents scale as $c^{(2n)}$ for Yang Mills scale with $N_c$ as $c^{(2n)} \sim N_c^{2-2n} $ which implies that for all 
\begin{equation}
c^{(2n)} \sim N_c^{2-2n}\; .
\label{cscale} \end{equation}  
Thus, as $n$ increases the $c^{(2n)}$ decrease parametrically quite quickly with $N_c$.  For example for $n=5$, this suppression factor is $N_c^{-8}$ which for $N_c=3$ is $3^{-8} \approx .0001524$.  The principal difficulty this poses is that small numbers are often hard to compute accurately via numerical means and are particularly difficult to extract accurately via Monte Carlo methods.  Thus, it may be hard to compute $c_n$s  up to even moderately large $n$ and those that are computed are apt to be noisy.  This is likely to limit range in $n$ that one can explore.   Moreover, the noise in the $c_n$ will translate to noise in the $f_n$.  Note that the scaling in Eq.~(\ref{cscale}) implies that all of the $f_n \sim N_c^{-2}$  and, thus, are parametrically small.  This means that the growth of $f_n$ with $n$ will also be parametrically small.   Detecting clear signs of growth of $f_n$ with $n$ might be difficult to see without going to larger $n$ given both the parametrically small growth and the substantial noise in the extraction of the $f_n$.  

Fortunately,  the numerical difficulties due this this $1/N_c$ suppression are likely to be  much less severe then the ones associated with chiral suppression for models satisfying conditions \ref{cond1}-\ref{cond9}.   In particular,   there is no reason to believe that  $N_c^{-2}$ suppression in the growing value of $f_n$ should be masked at small $n$ by a leading-order contribution as happens with chiral suppression.  Thus,  there is no reason to suspect that $1/N_c$ suppression would  induce the type of the behavior seen in Fig.~\ref{fig:toy} where the evidence of asymptotic growth is undetectable up to some comparatively large value of $n$ where it sets in suddenly.  Assuming this to be true, this greatly improves the prospects of seeing the effect in pure gauge theory than in QCD. 

One strategy to reducing  numerical difficulties associated with $N_c^{-2}$ suppression is to reduce $N_c$.  Rather than studying SU(3) pure gauge theory, one could study SU(2).  There are two virtues to doing this: the calculations intrinsically require fewer computational resources possibly allowing computations of $f_n$  with more accuracy or to larger $n$, and the $N_c^{-2}$ suppression effect for each $f_n$  are reduced simplifying the task of identifying growth of $f_n$.  Of course, there is also a downside---SU(2) Yang-Mills theory is obviously further from QCD then SU(3) Yang-Mills theory.  Never-the-less if one found evidence that SU(2) pure gauge theory appeared to be consistent with conditions \ref{cond1}-\ref{cond6}, it would certainly make a QCD-based solution to the strong CP problem far more plausible.

There is a hierarchy of theories starting with QCD itself, going through QCD with artificially large quark masses and SU(3)  pure gauge theory to SU(2) pure gauge theory.   As one descends this hierarchy, the theories become progressively less like QCD, but if the $f_n$ coefficients grow with $n$ asymptotically, then evidence for it becomes progressively more accessible.  It is noteworthy however, that at present, even toward the bottom of the hierarchy, too few $c_n$ coefficients have been computed to even begin exploring the issue---the state of the art for Yang-Mills theory\cite{Bonati:2015sqt}, only goes up to calculations of $c^{(6)}$ allowing only two $f_n$s and the $c^{(6)}$ computations are essentially bounds rather than well determined values.  Thus, at present, for  one can compute at most $f_1$ and  bound $f_2$ for Yang-Mills theories; this is insufficient to see even the hint of a trend valid at large $n$.  However, it is at least conceivable with a natural increase computer power and algorithmic development, it may be possible to extract a couple of more $f_n$s in the not-too-distant future. Even if these showed an increase with $n$, it would be insufficient to establish compelling evidence for conditions \ref{cond1}-\ref{cond6}.  However, it could be  suggestive of it.  Thus, in the short term evidence of this sort will necessarily be quite indirect.

There is another way to proceed, namely the study quantum field theories that may be quite far removed from QCD (including theories in space-time dimensions smaller than 3+1).  As noted previously, a principal reason to be skeptical that QCD could satisfy conditions \ref{cond1}-\ref{cond9}, is that over decades of experience with quantum field theory, the community has never encountered a theory that satisfies anything  resembling these conditions.  The converse, however, is also true.  Were one to find {\it any} quantum field theory where there is a reliable way to show that an analog of conditions \ref{cond1}-\ref{cond9} (or even conditions \ref{cond1}-\ref{cond6}) can be shown to hold and to emerge naturally from the structure of the theory, it would greatly increase the plausibility of this class of solution to the strong CP problem.  Moreover, if one were to find a theory that did satisfy the analog of these conditions, there is the prospect that one might well gain insight into the underlying mechanism causing the conditions to be satisfied and to develop a physical picture of what is behind it.  This in turn, might hint at what could be happening in QCD.  

This prospect suggests the possible utility of a review of known quantum field theories (in any number of dimensions) that are  tractable in some manner and have some analog to the $\theta$-term.  Such theories could be tractable for a number of reasons.  They could be one of the rare examples of an exactly solvable model.  Alternatively they  could be tractable numerically on a lattice in a manner that allowed testing the conditions.  Clearly numerical treatments on a lattice of theories with  nontrivial  topology are far more likely to be viable for theories in lower space-time dimensions than in 3+1 dimension.  Consider, for example, ref.~\cite{Fukaya:2003ph}, a calculation of a theory in 1+1 dimensions that was viable more than a decade and half ago but an analogous 3+1 dimensional calculation would be impractical today.   A third class of tractable examples would be a theory that is known to be  solvable in some limit and that admits systematic corrections away from that limit.  Of course with this third class, one must bear in mind that any conclusions based on  such a study may give misleading results (after all the entire premise of this paper is that the chiral expansion could give misleading results for $\theta$ dependence).   It is conceivable that one could identify one or more such theories could satisfy some type of analog of conditions \ref{cond1}-\ref{cond6}) that may not have been recognized as such in the past.

In searching for such a theory, the analogy to QCD need not be perfect.  Of course, the further the theory is away from QCD, the less evidence for the analog of conditions \ref{cond1}-\ref{cond6} suggests that QCD will also satisfy these.  Similarly, the weaker the analogy of the quantities in the tractable theory to those in QCD, the weaker the suggestion that QCD will satisfy them.   Never-the-less, if scenarios that solve the strong CP without invoking BSM physics in the manner considered in this paper are correct, then in the short term the best prospect for finding evidence for them---albeit rather indirect evidence---is via the study of tractable analog theories.

The author gratefully acknowledges the support of the US Department of Energy.  Comments by A. Cherman were extremely helpful.

\bibliography{StrongCP_QCD_chiral_prd}
\end{document}